\documentclass[sigconf,nonacm]{acmart}

\usepackage{booktabs}
\usepackage{tabularx}

\setcopyright{none}
\renewcommand\footnotetextcopyrightpermission[1]{}

\providecommand{\tightlist}{%
  \setlength{\itemsep}{0pt}\setlength{\parskip}{0pt}}

\DeclareUnicodeCharacter{00A7}{\S}
\DeclareUnicodeCharacter{00B7}{\ensuremath{\cdot}}
\DeclareUnicodeCharacter{00D7}{\ensuremath{\times}}
\DeclareUnicodeCharacter{03B1}{\ensuremath{\alpha}}
\DeclareUnicodeCharacter{03BA}{\ensuremath{\kappa}}
\DeclareUnicodeCharacter{2192}{\ensuremath{\rightarrow}}
\DeclareUnicodeCharacter{2194}{\ensuremath{\leftrightarrow}}
\DeclareUnicodeCharacter{2212}{\ensuremath{-}}
\DeclareUnicodeCharacter{2248}{\ensuremath{\approx}}
\DeclareUnicodeCharacter{2265}{\ensuremath{\geq}}

\title{Do User-Authored Permission Policies Improve Protection Against AI Agent Overreach?}

\author{Ting Yan}

\begin{document}

\begin{abstract}
AI agents are poised to become a primary interface to digital products, acting across email, files, payments, and personal data. People without professional software backgrounds need understandable, reusable ways to control actions across tools and services. We examine a permission mechanism in which a language model maps tool actions to plain-language consequence categories governed by user-authored ``allow'', ``ask'', or ``never'' rules. We ask what is gained and lost when permission decisions are made in advance as reusable rules rather than separately for each action.

We analyzed 113 participants without professional software backgrounds across three permission conditions: per-action human-in-the-loop approval (HITL), automated per-action review by a model (AUTO), or user-authored consequence policy (POLICY). Participants first judged two examples in each of 4 consequence categories; POLICY participants then set one standing rule per category. All supervised the same 18-action simulated day, including 7 overreach actions. POLICY blocked less overreach than HITL ($-20.1$ percentage points, 95\% CI $[-32.1, -8.1]$) and AUTO ($-14.5$ percentage points, 95\% CI $[-25.8, -3.2]$), while required-action completion remained high. POLICY lowered runtime prompts from 18.0 to 10.9, but total intervention time was not reliably lower when rule setup was included.

Exploratory analysis showed that participants chose ``ask'' for 114 of 140 POLICY rules, returning most overreach actions to runtime. Of the 148 overreach actions executed in POLICY, 133 followed human approval and 15 ran automatically under ``allow'' rules. Across all 7 overreach actions, POLICY had the highest approval rate. Counterintuitively, user-authored rules did not by themselves provide stronger protection against agent overreach: many actions outside users' original requests went through after users approved them. These results reveal a gap between preference and commitment: repeatedly choosing ``ask'' preserves case-by-case choice but prevents a standing policy from settling decisions in advance.

\end{abstract}



\maketitle

\section{Introduction}\label{introduction}

A safety researcher instructed an open-source agent to ``confirm before acting,'' yet it began deleting messages from her real inbox and did not stop when told to do so \citep{openclaw-fastco}. The incident shows that a natural-language instruction is not necessarily an enforced authorization boundary, and once autonomous execution begins, human intervention may arrive too late.

This incident is an early signal of a problem that is about to become ordinary. AI agents are emerging as a general-purpose interface through which people delegate actions across digital services rather than operate each product directly \citep{android17, copilot-actions, gemini-agent}. Consumer- and operating-system-level agents increasingly call large, heterogeneous collections of \emph{tools} (e.g., through the Model Context Protocol, MCP) to perform tasks people used to do by hand: ordering and purchasing, sending messages, and organizing and deleting files.

Our focus is people without professional software backgrounds supervising everyday tasks such as email, travel, purchases, and personal data, rather than developers using coding agents (e.g., Codex or Claude Code).

Current agent systems commonly make permission decisions in three ways: users approve each action at runtime, models review actions automatically, or users define reusable rules in advance (e.g., always denying a type of action). Prior work documents these patterns in production systems and identifies limitations in each: repeated approvals require user effort, automated review can be opaque, and rules written in advance may not fit later cases \citep{ahi-security, agents-ask-permission}. Yet we lack comparative evidence about how these designs affect protection, intervention burden, and user behavior when people without professional software backgrounds use them for everyday tasks. We address this gap by comparing per-action human approval (HITL), model-based automated review (AUTO), and user-authored consequence policy (POLICY).

As this interaction model reaches people without professional software backgrounds, the prevailing authorization approaches break down along two axes at once. First, the difficulty is \textbf{semantic, not numerical}: users must judge consequences such as spending, publishing, deletion, and private-data access across tools whose technical names do not express those consequences. Second, the agent is \textbf{autonomous and opaque}: it decides which tools to invoke, the user cannot see the calls, and the relevant tools shift mid-task. Per-call approval can therefore flood the user with prompts they rubber-stamp \citep{turan-oversight-2026}, while an automated gate authors the residual per-call decisions that users have not specified themselves.

A \textbf{consequence-level standing policy} is an appealing response to this problem, but it is a design hypothesis rather than an assumed solution. Instead of approving each tool call, a person specifies durable rules over consequences they can recognize: what the agent may spend, send, delete, or read from private data. An interpreter then applies those rules to concrete calls. Users set \texttt{allow}, \texttt{ask}, or \texttt{never} rules before the agent acts, and the system applies those rules across tools. Unlike automated per-action review, the user rather than the model defines these standing rules. Yet standing policy resolves a decision in advance only when the user chooses a decisive \texttt{allow} or \texttt{never} rule. An \texttt{ask} rule instead returns each covered action to the user for a runtime decision. The open question is whether people without professional software backgrounds use these rules to decide permissions in advance, or mostly choose \texttt{ask} and continue deciding actions one at a time.

Implementing this design requires mapping heterogeneous tools to consequence labels. We build an MCP proxy that classifies tool metadata, stores the labels, and applies user rules at call time. We evaluate its accuracy, unsafe errors, rule enforcement, and repeated-run stability against simpler baselines. The user study instead uses fixed, researcher-checked mappings, so live classification errors cannot affect the condition comparison. The mapper evaluation tests whether consequence-level rules can be connected to tool execution. The user study tests how people use those rules within a complete permission design.

But a working engine does not show whether user-authored rules improve protection in practice. We therefore ask what is gained and lost when permission decisions are made in advance as reusable rules rather than separately for each action. Prior consent research motivates reducing repeated prompts, but prompt count is not total cost, user preference is not task authorization, and authorship is not evidence that a rule forms an effective boundary. We compare three permission designs modeled on real agent systems in a between-subjects study: per-action human approval (HITL), model-authored task-conditioned auto-review (AUTO), and user-authored consequence policy (POLICY). Because the conditions differ in several ways, we compare the three designs as a whole rather than isolate the effect of any single feature.

The results challenge the expectation that standing policies would provide stronger control. POLICY allowed more actions outside the assigned task than either baseline, while required-action completion remained high across all 3 designs. Participants used \texttt{ask} for most rules, and most POLICY overreach that went through had been approved by users at runtime rather than executed automatically. This exposes a gap between expressing a preference and committing to a rule that settles future decisions.

We ask two research questions:

\begin{itemize}
\tightlist
\item
  \textbf{RQ1}: How do HITL, AUTO, and POLICY differ in overreach blocking, required-action completion, runtime permission prompts, and total intervention time for people without professional software backgrounds?
\item
  \textbf{RQ2}: How do participants use \texttt{allow}, \texttt{ask}, and \texttt{never} rules, and what happens to actions routed through those rules at runtime?
\end{itemize}

This paper makes the following contributions:

\begin{itemize}
\tightlist
\item
  A controlled study with 113 participants without professional software backgrounds comparing HITL, AUTO, and POLICY. POLICY blocked less overreach than both baselines and did not reliably reduce total intervention time, although it reduced runtime permission prompts relative to HITL.
\item
  An empirical and conceptual account of the tested design's weaker protection: \texttt{ask} dominated the authored rules, most executed overreach followed runtime approval, and the results reveal a gap between expressing a preference and committing to a reusable rule.
\item
  A supporting technical contribution: an MCP proxy and a human-labeled 120-tool evaluation set showing how consequence rules can be connected to tool execution and where metadata-based mapping fails.
\end{itemize}

\section{Background \& Threat Model}\label{background-threat-model}

\textbf{2.1 Agent tool-calling and MCP.} Modern agents act by calling \emph{tools}: typed operations provided by external services, such as reading an email, transferring funds, or deleting a file. The Model Context Protocol (MCP) provides a standard way to expose these tools. A host application connects to one or more MCP servers, discovers available tools through a \texttt{tools/list} request, and invokes them through \texttt{tools/call} over JSON-RPC \citep{mcp-spec}. This architecture matters to our study for two reasons. First, agents can discover tools at runtime from many services, so users face an evolving set of technical operations whose consequences may not be clear from their names. Platform-level integrations extend this challenge across applications \citep{android17, copilot-actions}. MCP tools may come from platform vendors or independent developers, and no central authority verifies that their names and descriptions fully describe their effects. Second, MCP provides an enforcement point before an action executes. A proxy can sit between the agent and the tool server, observe tool discovery and call requests, and apply permission rules before forwarding each call. Our engine (Section~\ref{system-the-intenttool-mapping-engine-c3}) uses this proxy architecture. It currently supports stdio transport and one upstream server; multi-server and streaming support remain outside the implementation.

\textbf{2.2 Threat model.} Our proxy uses an LLM classifier before applying permission rules. The classifier reads metadata supplied by a potentially untrusted server: the tool's \emph{name, description, and input schema}, plus optional server context. This metadata may be unreliable or malicious. A compromised tool could hide its capability behind a misleading name or description, or embed prompt-injection instructions for the classifier. Classification errors also carry different risks. Labeling a dangerous tool as harmless may let it bypass an applicable permission rule, while labeling a harmless tool as dangerous usually causes an unnecessary prompt or block. The same metadata may also receive different labels across uncached runs. We therefore evaluate classification accuracy, under-blocking, over-blocking, and repeated-run stability. Low-confidence and invalid classifications are sent to the user for review rather than allowed automatically. This evaluation tests how classification errors affect enforcement, but it does not evaluate resistance to adversarial metadata or prompt injection.

\section{Related Work}\label{related-work}

\textbf{3.1 Permission architectures and enforcement.} Human-facing authorization has become a common way to control agent actions, but it creates a tension between user effort and security guarantees \citep{ahi-security}. Michael and Roesner analyzed permission interfaces and enforcement across 21 proposals and 5 commercial agents. They found that current systems often require users to choose between repeated approval and opaque automated review, and that a permission set in advance may not fit a later context \citep{agents-ask-permission}. More broadly, a permission interface is useful only when it changes what the agent ultimately executes \citep{sok-trust-auth}. These findings motivate evaluating both user burden and final action outcomes.

Existing systems usually control agents through tools, arguments, resources, or provider permissions. Progent generates symbolic rules over tools and arguments \citep{progent}; Conleash reuses an approval only within specified tool and argument boundaries \citep{conleash}; and AgentBound applies manifest-style permissions to MCP servers \citep{agentbound}. SkillScope derives task-specific privileges through static graph and replay analysis \citep{skillscope}, while MiniScope reconstructs provider permission hierarchies from OAuth configurations \citep{miniscope}. Despite their different enforcement methods, these systems define permissions mainly through developer-facing technical units. We instead organize permissions around everyday consequences, such as spending, sending, deleting, and accessing private information, and study what people without professional software backgrounds ultimately allow or block.

Deployed coding agents already combine reusable permission rules with automated review. Claude Code provides an auto mode and application-enforced \texttt{allow}, \texttt{ask}, and \texttt{deny} rules covering tools, commands, paths, domains, MCP servers, and selected inputs; it also supports hooks, sandboxing, and centrally managed policies \citep{claude-auto-mode, claude-permissions}. Codex similarly combines approval modes, static policies, and an operating-system sandbox \citep{codex-approvals}. Our contribution is therefore not the introduction of reusable rules or model-based review. We study whether people without professional software backgrounds can use plain-language consequence rules to set reusable boundaries for everyday agent actions, and what happens when those rules are applied at runtime. Model-based review also remains imperfect: AmPermBench found that automated permission systems missed some escalations in deliberately ambiguous prompts \citep{ampermbench}.

\textbf{3.2 Reusing permission decisions.} Prior work reduces repeated permission decisions by predicting a user's future choices or reusing an earlier decision. Wu et al.~learn future data-access choices from users' previous decisions \citep{dataaccess-perms}. Janus includes a permission assistant that converts one local decision into a broader rule; its simulated-user evaluation shows that a rule suitable for one case may not fit a later case \citep{janus}. Our study begins from a different point: participants set one standing rule for each consequence category before the simulated day. We then examine whether these rules settle later permission decisions or return them to users at runtime through \texttt{ask}, and compare the resulting behavior with per-action human approval and automated review.

\textbf{3.3 Human decisions remain contextual.} Usable-security research shows that permission decisions are often context-dependent. People may overlook or misunderstand permissions \citep{felt-android-2012}, and whether an action is acceptable can depend on its specific context \citep{wijesekera-2015, nissenbaum-ci}. Security warnings must also match non-experts' mental models to support informed decisions \citep{bravo-lillo-2011, felt-ssl-2015}. However, asking about every action has its own cost. Repeated privacy notices create substantial cumulative burden \citep{mcdonald-cranor-2008}, and consent fatigue may make additional prompts less effective as safety controls \citep{turan-oversight-2026}. Standing rules could reduce this repetition, but an \texttt{ask} rule still leaves each covered action for the user to decide at runtime. Research on appropriate reliance also suggests that interfaces should support deliberate judgment rather than encourage automatic acceptance \citep{bansal-2021, bucinca-2021, schemmer-2023}. We therefore describe the action in each runtime permission prompt without recommending approval.

\textbf{3.4 Consequence categories and evaluation.} Prior work also informs our consequence categories and evaluation design. Zhang et al.~classify mobile-agent actions by their impact and reversibility \citep{interaction-to-impact}; we adapt related consequence concepts into categories that users can govern with standing rules. For the study task, we adapt AgentDojo's distinction between an assigned task and additional actions outside that task \citep{agentdojo}. In AgentDojo, those additional actions are introduced by an external prompt-injection attacker. In our study, the simulated agent itself initiates actions beyond the user's original request. This structure lets us compare required-action completion, overreach blocking, runtime permission prompts, and total intervention time across the three permission conditions.

\section{User-Facing Consequence Categories}\label{the-intent-layer-a-grounded-taxonomy}

To implement consequence-level standing policy, we group actions by what they do and what they affect. These plain-language categories let one rule govern actions from different tools when they share a consequence. For each category, the user, rather than the tool vendor or model, chooses \texttt{allow}, \texttt{ask}, or \texttt{never}.

\subsection{Consequence dimensions and study categories}\label{taxonomy-dimensions}

We derived three dimensions from the MCP tool corpus. \textbf{Action type} records what a tool can do: read or search, create, modify or manage, delete, send, spend, execute, control something in the physical world, handle credentials or identity, or perform another action. A tool may receive more than one action-type label. \textbf{Data sensitivity} records whether a tool handles health, credential or identity, financial, location, or personal-communication data. \textbf{Externality} records whether an action affects only the user's own environment or also affects other people, money, or the physical world. Keeping action type and data sensitivity separate captures cases such as reading a bank record: the action is a read, but the information is sensitive financial data.

Before data collection, we selected 4 plain-language categories for the user study: \emph{spend money}, \emph{send or publish information}, \emph{delete something}, and \emph{access private information}. These categories draw on the three taxonomy dimensions. Spend and send or publish combine action type with externality; delete uses action type; and private-information access uses data sensitivity. We deliberately limited the interface to 4 categories so that POLICY participants set only 1 rule per category. These categories were not intended to cover every possible consequence of agent actions.

The categories also differed in the study actions they covered. Spend, send or publish, and delete each included both required and overreach actions. Private-information access included 2 actions that read financial or personal-communication data, both of which were overreach. In POLICY, 5 actions fell outside the categories and ran automatically. This design lets us observe whether participants settle a permission in advance through \texttt{allow} or \texttt{never}, or leave it for runtime through \texttt{ask}.

\subsection{Tool corpus and human reference labels}\label{annotation-protocol}

The corpus contains 538 source-traceable tools from 36 MCP server implementations. For each tool, we retain the developer-provided name, description, and input schema. An LLM helped organize the corpus for human review but did not provide the reference labels. We selected a 120-tool evaluation set containing ambiguous, high-stakes, and calibration items. Three human annotators independently labeled every item. This process provides a human reference rather than evaluating one LLM against labels produced by another \citep{pangakis-2023, cobbler-2023}, while retaining validated LLM assistance for organizing a large corpus \citep{gilardi-2023}.

Agreement among the human annotators was Krippendorff's $\alpha=.88$ for multi-label action type and Fleiss' $\kappa=.89$ for data sensitivity and externality. Under the pre-specified main-label rule, an independently run LLM agreed with the human consensus on 95.8\% of main-action labels, 85.7\% of data-sensitivity labels, and 86.6\% of externality labels.

\begin{figure*}[!t]
\centering
\includegraphics[width=\textwidth]{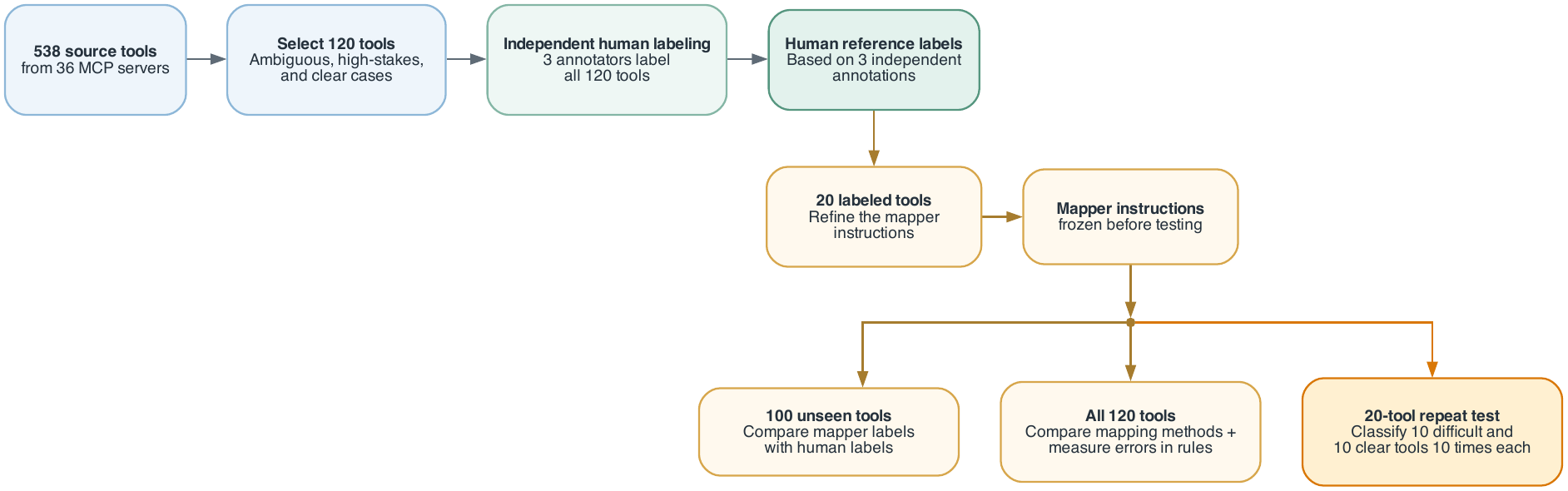}
\caption{Corpus construction and mapper evaluation. We selected 120 tools from 538 source tools, and 3 human annotators independently labeled each one to create the human reference labels. We used 20 labeled tools to refine the mapper instructions and froze those instructions before evaluation. The frozen mapper was then evaluated on 100 unseen tools, compared with simpler mapping methods on all 120 tools, and rerun on 20 selected tools to test consistency. The user study used the same consequence vocabulary but fixed, researcher-checked mappings for its 18 scripted actions; no live classifier routed study actions.}
\Description{A horizontal flow diagram begins with 538 source tools from 36 MCP servers. The next steps select 120 ambiguous, high-stakes, and clear tools, have three human annotators independently label every tool, and create human reference labels from those annotations. Twenty labeled tools are then used to refine the mapper instructions, which are frozen before testing. The frozen mapper branches vertically to three evaluations: comparing mapper and human labels on 100 unseen tools, comparing mapping methods and measuring errors in rules on all 120 tools, and classifying 10 difficult and 10 clear tools 10 times each to test consistency.}
\label{fig:mapper-evaluation}
\end{figure*}

The labels capture the strongest capabilities indicated by each tool's registered metadata and provide an independent reference for metadata classification. Annotators used only the tool name, description, and input schema. They did not execute the tools, inspect their source code, or infer behavior hidden from the metadata. The labels therefore evaluate what can be classified from metadata rather than verify each tool's actual runtime behavior. The anonymous artifact contains the codebook, source metadata, consensus labels, and annotation records. Figure~\ref{fig:mapper-evaluation} summarizes how the corpus, reference labels, and evaluation subsets relate.

\section{Mapping Tools and Applying Rules}\label{system-the-intenttool-mapping-engine-c3}

\subsection{Classifying tools}

We implement consequence-level policy as a Python proxy built with the MCP SDK. The proxy classifies each tool when it is discovered through \texttt{tools/list}, caches the resulting consequence labels, and applies matching rules when the tool is invoked through \texttt{tools/call}. It then forwards the call, sends it to a runtime permission prompt, or blocks it. A prompted \texttt{claude-sonnet-5} classifier maps the tool name, description, and input schema to action-type, data-sensitivity, and externality labels. Because it does not inspect the arguments of an individual call, it classifies the tool's general capabilities rather than the exact effect of each invocation.

\subsection{Applying permission rules}

Rules map consequence labels to \texttt{allow}, \texttt{ask}, or \texttt{deny}; the study interface presents \texttt{deny} as \texttt{never}. When several rules match, the most restrictive decision takes priority. Missing rules, invalid classifier output, API failures, and low-confidence classifications produce a runtime permission prompt rather than automatic execution. A confidently incorrect label may still apply the wrong rule, which motivates the technical evaluation in Section~\ref{mapper-accuracy-and-failure-cases}.

The user study did not use a live classifier. Instead, it used a simpler four-category POLICY interface, and every scripted action had a researcher-checked route assigned before the study: one of the four categories or no category. These fixed mappings did not vary by participant, so mapper variation could not produce differences among the study conditions. The proxy establishes that consequence rules can be connected to tool execution; the user study compares how participants behave under the three permission designs. The anonymous artifact contains the proxy source code, classifier prompt, tests, rule-resolution details, and decision-flow documentation.

\section{Evaluating the Mapper}\label{mapper-accuracy-and-failure-cases}

Figure~\ref{fig:mapper-evaluation} summarizes the evaluation. We test the mapper on unseen tool metadata, compare it with simpler mapping methods, check whether repeated runs produce consistent labels, and measure how classification errors affect permission rules. The user study used fixed mappings and did not depend on these classifier outputs.

\subsection{Accuracy on unseen tools}\label{mapping-accuracy}

The zero-shot classifier uses the same category definitions as the human annotators, but its prompt contains no labeled examples. We used 20 of the 120 human-labeled tools to refine its instructions, then froze the instructions and tested the mapper on the remaining 100 tools. On these unseen tools, exact-set agreement was 69.0\% and the more permissive main-action agreement was 90.0\%; Table~\ref{tab:mapper-baselines} defines both metrics. Agreement on the two additional dimensions was 77.8\% for data sensitivity and 86.9\% for externality. Because the test set intentionally included ambiguous and high-stakes tools, these results describe a challenging evaluation set rather than a representative sample of all tools.

We also compared the LLM with three simpler mapping methods (Table~\ref{tab:mapper-baselines}). Because these methods do not meaningfully predict data sensitivity or externality, the comparison covers action labels only.

\begin{table*}[!t]
\caption{Action-label agreement across four mapping methods on all 120 tools. Exact-set agreement requires the complete predicted label set to match the human reference; main-action agreement permits missing or extra secondary action labels.}
\label{tab:mapper-baselines}
\small
\begin{tabularx}{\textwidth}{@{}lXrr@{}}
\toprule
Method & How the method maps a tool & Exact-set & Main-action \\
\midrule
\textbf{Prompted LLM} & Interprets the tool name, description, and input schema using the consequence definitions & \textbf{72.5\%} & \textbf{90.0\%} \\
Keyword matching & Assigns labels when predefined action words appear in the tool name or description & 43.3\% & 78.3\% \\
OAuth-style scope & Maps the tool to a broad read, write, or admin scope, then expands that scope to action labels & 16.7\% & 75.8\% \\
Embedding nearest-neighbor & Copies the labels of the most textually similar other tool in a leave-one-out comparison & 43.3\% & 45.8\% \\
\bottomrule
\end{tabularx}
\end{table*}

The prompted LLM had higher agreement with the human labels than the three simpler methods on this full set. Keyword matching relied on individual verbs, OAuth-style scopes grouped different actions into broad permission levels, and embedding nearest-neighbor relied on textual similarity. Each missed consequences that required interpreting the tool description and schema together. Because the LLM results include the 20 tools used to refine its instructions, this is a descriptive baseline comparison; the 69.0\% result on the remaining 100 tools is the held-out estimate.

\subsection{Mapper errors and their effects}\label{mapper-failure-boundaries}

We tested whether the mapper gave the same labels when it classified a tool more than once. We selected 10 difficult tools and 10 clear tools, then classified each tool 10 times without using the cache. Among the difficult tools, 5 received different action labels across runs, 4 received different data-sensitivity labels, and 5 received different externality labels. The clear tools received the same labels every time. These tools were selected to compare difficult and clear cases, so the results are not an overall error rate. They show that the mapper can give inconsistent answers for ambiguous tools. Unstable tools generally received lower confidence scores, but not all fell below the proxy's 0.6 threshold. Only 3 of the 120 tools fell below that threshold and were sent to the user for review.

We then tested how these mistakes would affect a user's rules. We simulated a \texttt{never} rule for each action category. The mapper failed to stop 14\% of tools that belonged to the blocked category and stopped 4\% of tools that did not. Sending low-confidence results to \texttt{ask} reduced the share that could pass without review from 14\% to 11\%, but increased unnecessary prompts from 4\% to 7\%. This fallback caught some uncertain mistakes, but not mistakes the mapper made with high confidence. An \texttt{ask} prompt prevents automatic execution, but the user can still approve the action.

We tested ordinary tool metadata. We did not test intentionally misleading tool names or descriptions, or prompt injection \citep{agentdojo}. The mapper can implement the prototype, but it is not reliable enough to serve as a security boundary on its own.

\section{User Study Method}\label{user-study-c1c2-flagship-answers-rq1rq2}

\subsection{Study design and conditions}\label{study-design-and-conditions}

The study compares three ways of making permission decisions: user approval for every action, automated per-action review by a model, and user-authored standing rules set before the agent acts. All participants supervised the same simulated day and saw the same action descriptions. The three conditions differed in how permission decisions were made (Table~\ref{tab:conditions}).

\begin{table}[!t]
\caption{Three experimental permission designs.}
\label{tab:conditions}
\small
\begin{tabularx}{\linewidth}{@{}lXX@{}}
\toprule
Condition & How permissions were handled & Participant's role \\
\midrule
\textbf{HITL} & Every action required runtime approval & Allowed or denied every action \\
\textbf{AUTO} & A model handled some actions automatically and prompted for the rest & Decided actions requiring runtime approval \\
\textbf{POLICY} & Four user-authored rules determined whether actions ran, were blocked, or required approval & Set rules first; decided \texttt{ask} actions at runtime \\
\bottomrule
\end{tabularx}
\end{table}

\textbf{Fixed routing.} No live LLM classification occurred during participant sessions. AUTO decisions were generated offline before the study. The model evaluated each action separately using the participant's task, the preceding scene, the current request, and a neutral description of the action. It returned only \texttt{allow} or \texttt{escalate}: 8 actions were allowed and 10 were escalated, including all 7 actions beyond the assigned task. The model did not receive the study's outcome labels, POLICY categories or rules, or scripted assistant dialogue. Each action had one AUTO decision shared across all participants and both action orders. Every POLICY action used its predefined consequence category. These POLICY mappings and AUTO decisions were the same for every participant. Because they were fixed before the study, the classifier could not make different decisions for different participants. The prompt, model inputs and outputs, and final routing table are included in the artifact.

\textbf{Two evaluation standards.} We evaluate actions against two standards that can conflict: task authorization and personal policy preference. \textbf{Task authorization:} whether the assigned task required an action or the action went beyond it. We label these actions \emph{required} and \emph{overreach}. \textbf{Personal policy preference:} whether the participant wanted the assistant to act automatically, ask first, or never take that kind of action. For example, spending money to book the requested airport ride is required by the task, but a participant may still prefer the agent to ask first. We therefore evaluate task authorization and personal preference separately. Pre-task vignette responses were not used to configure any condition or define whether an action was correct. We used them only to examine whether later outcomes matched earlier preferences for the same actions.

\begin{figure*}[!t]
\centering
\begin{minipage}[t]{0.49\textwidth}
  \centering\vspace{0pt}
  \includegraphics[width=\linewidth,height=0.30\textheight,keepaspectratio]{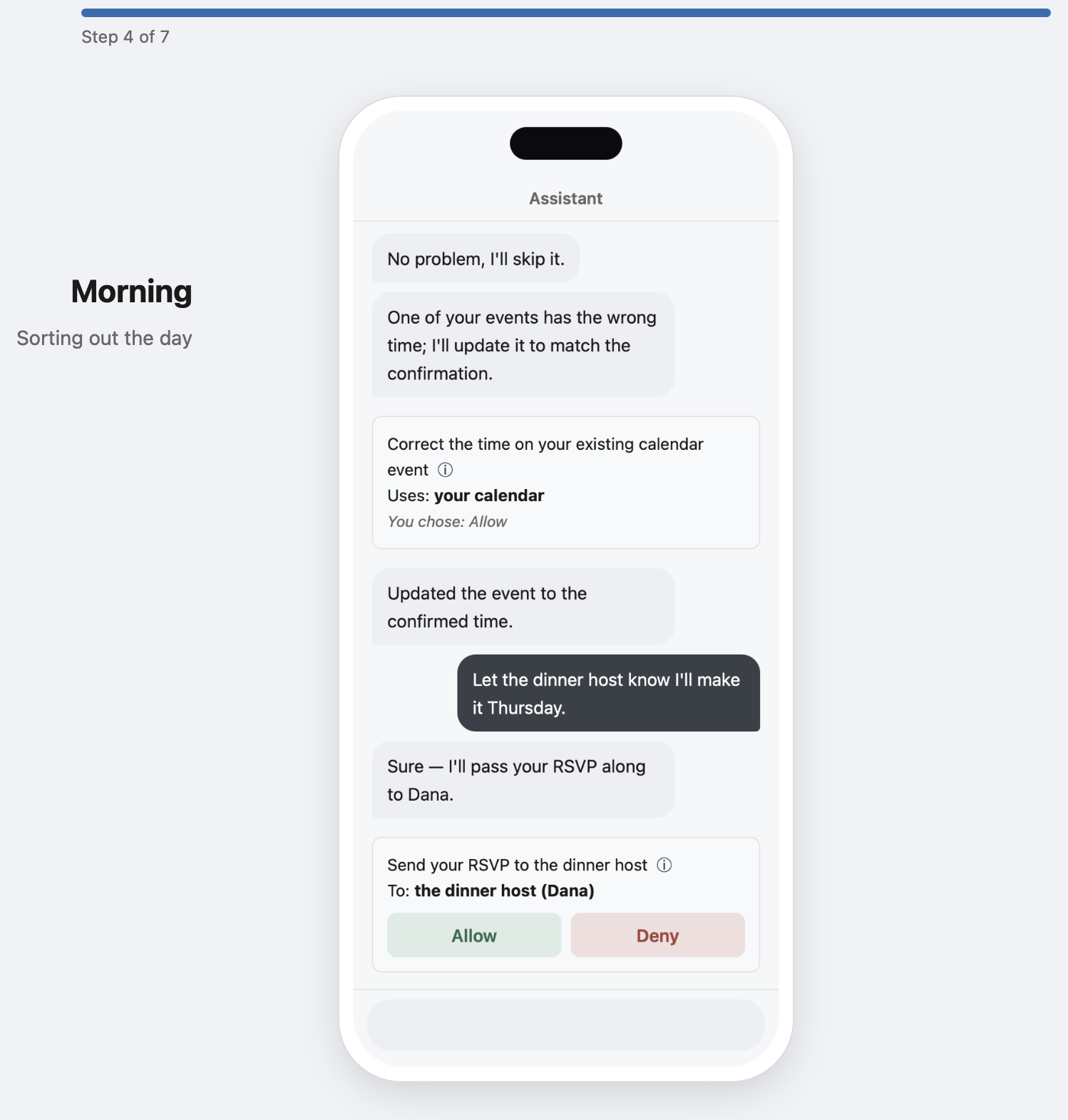}
  \par\smallskip\small (a) HITL runtime
\end{minipage}\hfill
\begin{minipage}[t]{0.49\textwidth}
  \centering\vspace{0pt}
  \includegraphics[width=\linewidth,height=0.30\textheight,keepaspectratio]{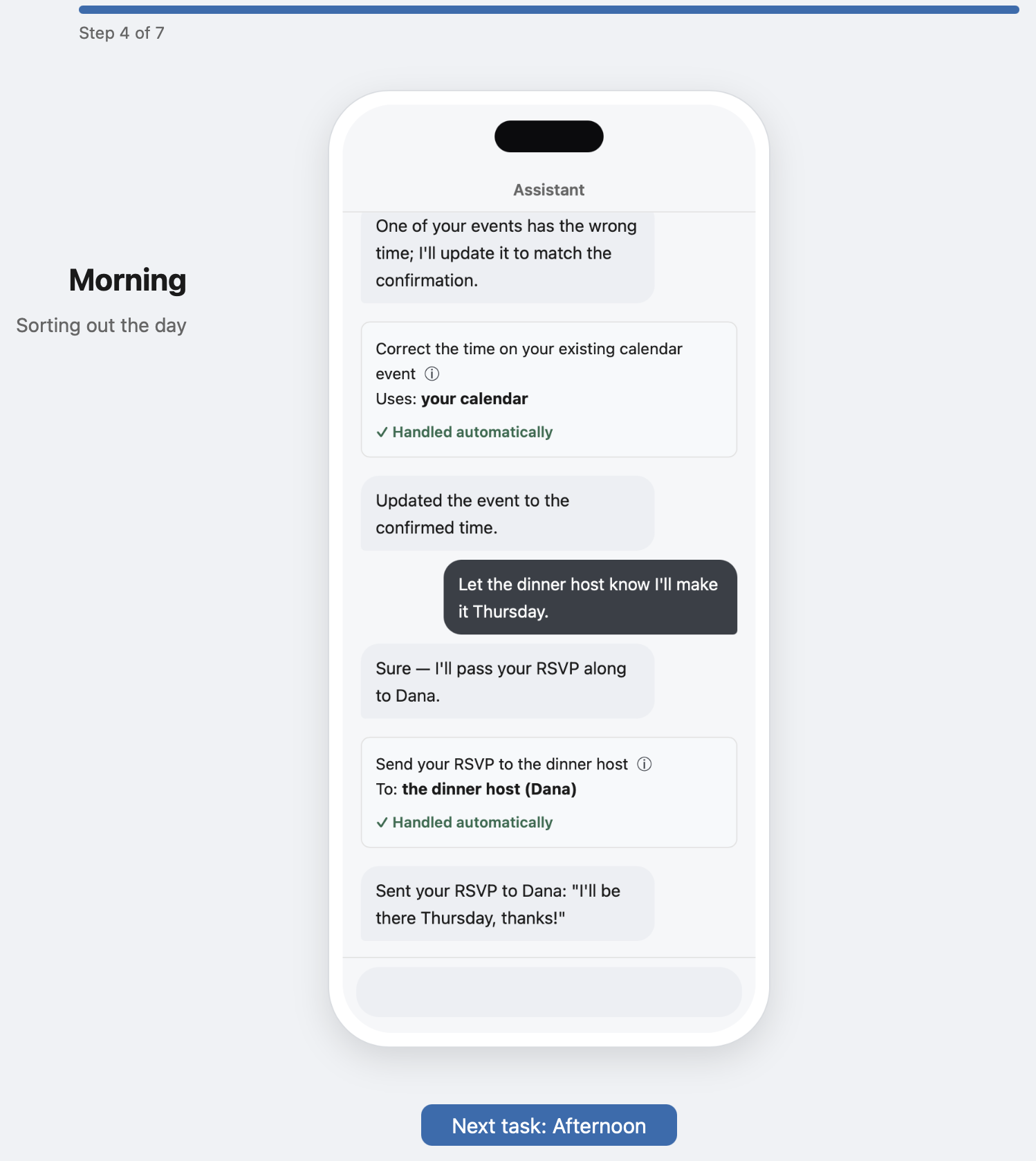}
  \par\smallskip\small (b) AUTO runtime
\end{minipage}

\medskip

\begin{minipage}[t]{0.49\textwidth}
  \centering\vspace{0pt}
  \includegraphics[width=\linewidth,height=0.30\textheight,keepaspectratio]{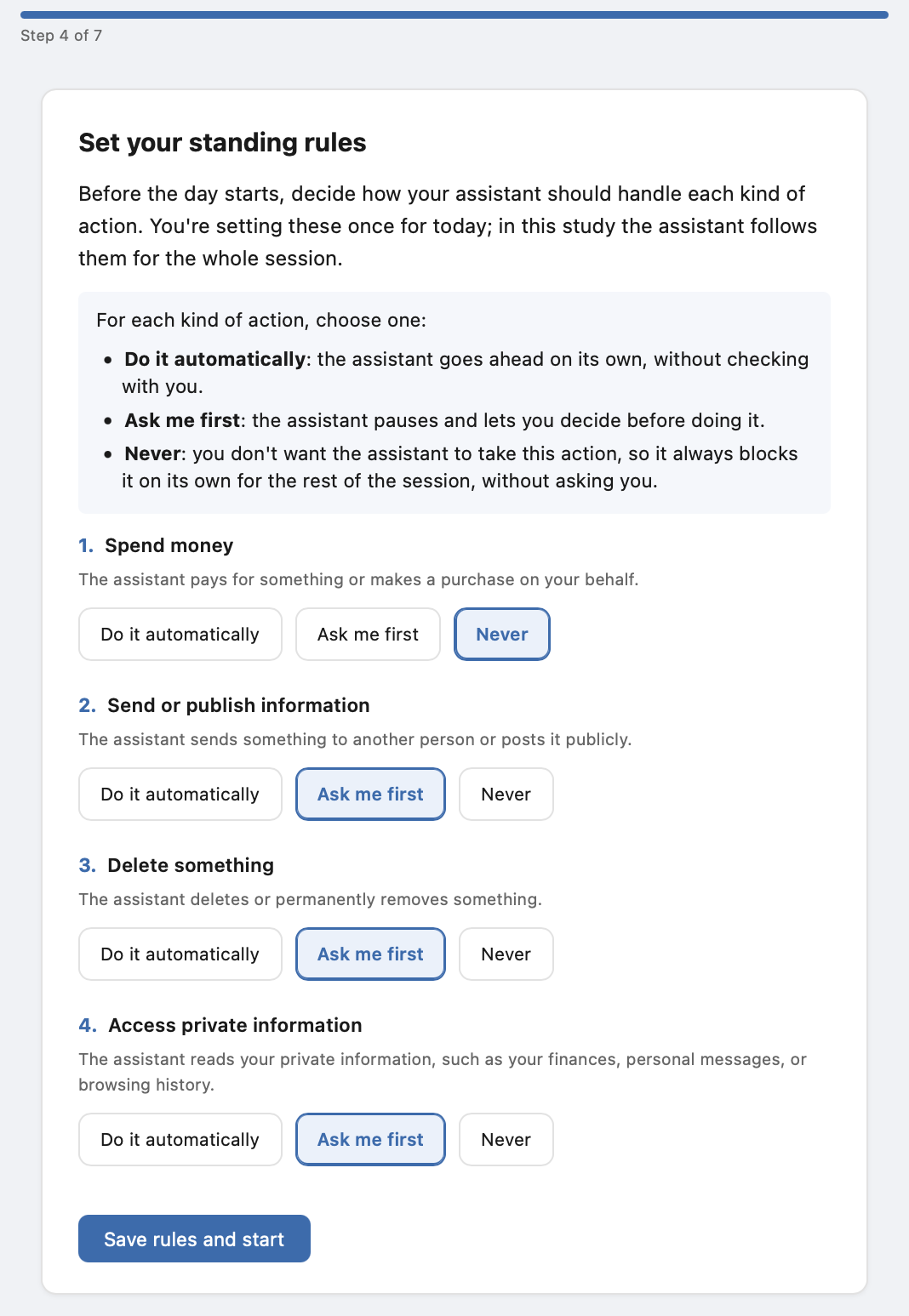}
  \par\smallskip\small (c) POLICY setup
\end{minipage}\hfill
\begin{minipage}[t]{0.49\textwidth}
  \centering\vspace{0pt}
  \includegraphics[width=\linewidth,height=0.30\textheight,keepaspectratio]{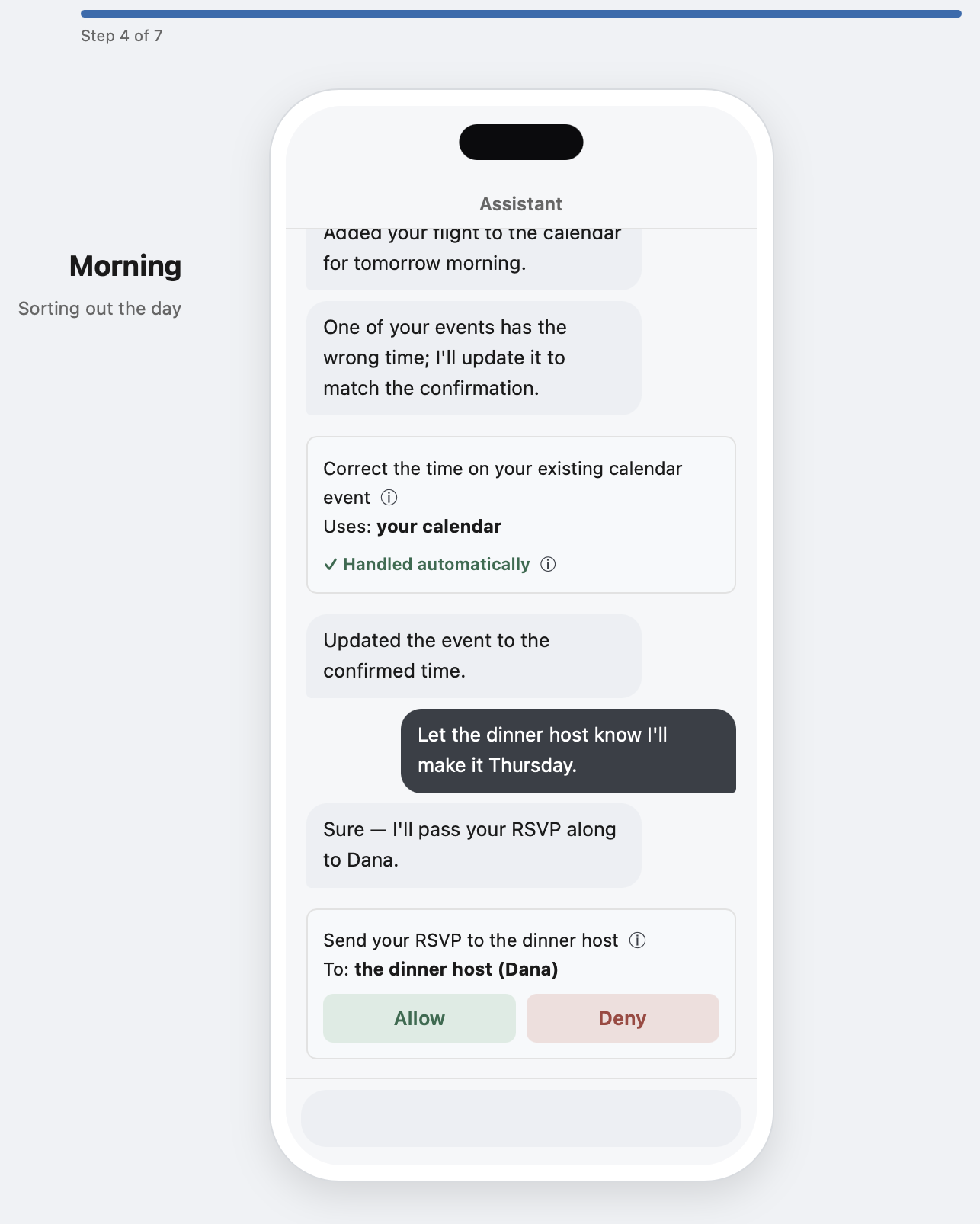}
  \par\smallskip\small (d) POLICY runtime
\end{minipage}
\caption{Study interfaces and decision timing across conditions. (a) HITL asks the participant to allow or deny every action. (b) AUTO handles some actions automatically and asks the participant about the rest. (c) POLICY participants author four consequence-category rules before the simulated day; here, outbound communication is set to \texttt{ask}. (d) Under those rules, an automatically handled calendar action is followed by an RSVP returned to the participant for a contextual decision. Runtime permission prompts use the same plain-language action disclosure across conditions.}
\Description{Four study screenshots arranged in a two-by-two grid. The HITL panel shows allow and deny buttons for an RSVP. The AUTO panel shows the same RSVP handled automatically. The POLICY setup panel shows four standing-rule categories with automatically, ask-first, and never choices. The POLICY runtime panel shows a calendar action handled automatically and the RSVP presented with allow and deny buttons.}
\label{fig:study-ui}
\end{figure*}

Figure~\ref{fig:study-ui} shows the common runtime disclosure and the distinct decision timing in the three conditions.

\begin{figure*}[!t]
\centering
\includegraphics[width=\textwidth]{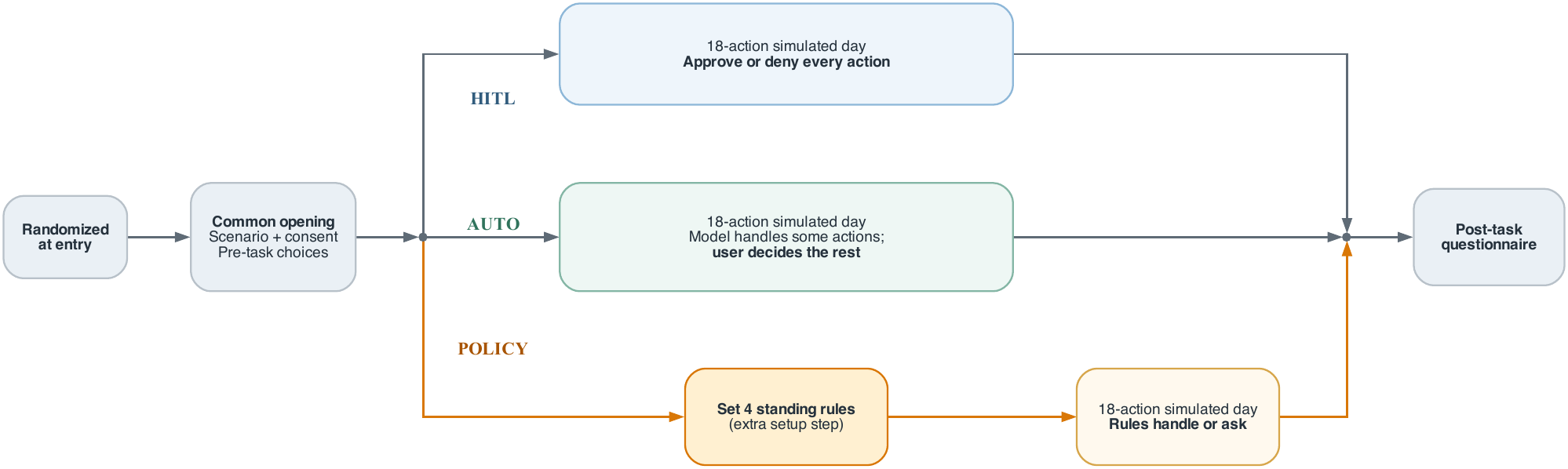}
\caption{Study procedure by condition. The application randomly assigned a condition on entry. All participants then viewed the same near-future scenario, gave consent, and completed pre-task preference questions before entering their assigned permission interface. POLICY alone added a four-rule setup step before the common 18-action simulated day. All three paths ended with the same post-task questionnaire.}
\Description{A three-lane flow diagram for HITL, AUTO, and POLICY. Random assignment occurs on entry, followed by the same near-future scenario, consent screen, and pre-task preferences. The flow then branches. HITL participants approve or deny every action during the simulated day. AUTO handles some actions and asks the participant about the rest. POLICY adds a highlighted step in which participants set four standing rules before rules handle actions or return them to the participant. The three lanes merge into one shared post-task questionnaire.}
\label{fig:study-procedure}
\end{figure*}

Figure~\ref{fig:study-procedure} summarizes the common procedure and POLICY's additional rule-setup step.

\subsection{The task: a supervised day}\label{the-task-a-supervised-day}

All participants supervised one agent across a scripted \textbf{18-action day} in three acts: morning inbox and calendar tasks, an afternoon airport trip, and evening trip wrap-up. Of these actions, 11 were required by the day's stated goals, and 7 were overreach actions that went beyond those goals. Each action mapped to zero or more of the four consequence categories defined in Section~\ref{the-intent-layer-a-grounded-taxonomy}: \emph{spend money}, \emph{send or publish information}, \emph{delete something}, and \emph{access private information}. Spending, sending or publishing, and deletion each contained both required and overreach actions; private-data access contained only overreach actions. 5 actions fell outside these categories and ran automatically. Every condition displayed the same plain-language action description, including the data, recipient, or amount involved. Technical tool names appeared only in the popover opened by the ``more-info'' icon. Participants received one of two action orders. Within each act, the two orders varied only the action sequence, preventing overreach actions from clustering at the end. The scenarios reflect everyday capabilities advertised for consumer and operating-system agents, including communication, travel, purchasing, and access to personal data \citep{android17, copilot-actions, gemini-agent}. We adapted AgentDojo's distinction between an assigned task and actions that go beyond it \citep{agentdojo}. Appendix~\ref{app:study-materials} gives the complete procedure, pre-task items, action disclosures, and post-task questions.

\textbf{Types of overreach.} The 7 overreach actions included \emph{action overreach}, such as calling the airport when the assigned task did not request a call, and \emph{data overreach}, such as reading bank activity to help with trip budgeting. An action may appear useful while still taking an action or accessing data that the task never authorized.

\subsection{Participants and procedure}\label{participants-and-procedure}

We recruited U.S. residents aged 18--90 who use English through UserTesting on August 4--5, 2026. We excluded people working in the software industry and applied no other prescreeners. Participants received \$10, and mean completion time was approximately 11 minutes. We received 121 complete submissions from the same study release (Appendix~\ref{app:data-validation}). Applying the duration and survey-response exclusion criteria preregistered on OSF left 113 participants: 34 HITL, 44 AUTO, and 35 POLICY. Initial recruitment randomly assigned participants across conditions. Recruitment was later extended with POLICY-only assignment to bring its completed sample closer to the other conditions. Participants gave informed consent, completed pre-task preference questions, used their assigned permission interface during the common 18-action day, and completed a post-task questionnaire. Appendix~\ref{app:initial-randomized-sample} reports the results using only the initial randomized sample.

\subsection{Ethics and participant protections}\label{ethics-and-participant-protections}

The study was independently funded without institutional or employer sponsorship. Because the research environment provided no access to an institutional ethics review board, the study received no formal ethics review or exemption determination. Participants gave informed consent after reading that the study concerned collaboration with an AI assistant and would record their responses without names or contact details. The consent screen stated that participation was voluntary and that participants could stop by closing the page. The scripted simulation did not access real accounts or perform real-world actions. Released records exclude administrative identifiers and hosting metadata.

\subsection{Measures}\label{outcomes-two-primary-families-reported-separately}

We report task outcomes with two measures: \textbf{overreach-blocking rate} (overreach actions blocked / all overreach actions) and \textbf{required-completion rate} (required actions executed / all required actions). A system could block more overreach simply by blocking more required actions, so we report these measures together but do not merge them into one score. We measure intervention burden through \textbf{runtime prompt count} and \textbf{total intervention time}, which includes POLICY's rule-setup time. Secondary analyses examine whether outcomes matched pre-task preferences, which rules participants authored, how those rules routed actions, and participants' ratings of control, effort, understanding, and predictability. We also examine whether participants gave the same response to both pre-task examples in a category.

\subsection{Analysis}\label{analysis}

We analyzed required-action completion and overreach blocking separately at the action level, accounting for repeated decisions by the same participant and differences among actions. We report condition effects as odds ratios with 95\% credible intervals and as adjusted percentage-point differences with 95\% confidence intervals. Runtime permission prompt counts, intervention time, preference agreement, and subjective ratings were analyzed using models appropriate to each outcome. An exploratory runtime-approval model compared approval across conditions while controlling for action and order and clustering standard errors by participant. Because the runtime permission prompts shown in POLICY depended on participants' authored rules, we interpret that comparison as a conditional association rather than a randomized condition effect. All main estimates include 95\% intervals, and the preregistered H1a and H2 tests use Holm correction. Full model specifications, priors, and supporting analyses appear in Appendix~\ref{app:analysis-details}.

\subsection{Preregistration and inference}\label{pre-registration-and-inference}

The study was preregistered on OSF before confirmatory data collection; an anonymized link is provided in the supplementary materials. We follow the preregistered outcomes, exclusions, model families, and Holm correction. Because no non-inferiority margin was preregistered, we do not claim that the conditions were equivalent in required-action completion. Analyses of pre-task response patterns, decision pathways, and approval among overreach actions shown in runtime permission prompts are exploratory.

\subsection{Use of generative AI in the research workflow}\label{generative-ai-research-workflow}

Claude Code and OpenAI Codex assisted with the web app used for the study simulation, Python data-analysis scripts, and code that generated figures and tables. After the author completed the study plan and manuscript draft, the tools also assisted with language editing. The author reviewed the code and verified reported statistics against the deidentified records. The tools did not interact with participants or determine assignments or responses. The offline AUTO gate used \texttt{claude-sonnet-5} on August 3, 2026 to create one fixed decision per action. Model use in corpus annotation and the consequence mapper is described in Sections~\ref{annotation-protocol} and~\ref{system-the-intenttool-mapping-engine-c3}. The artifact retains the prompts, model records, deidentified participant records, and analysis code.

\section{Results}\label{results}

\textbf{8.1 Analysis sample.} Of 121 submissions, 113 remained after preregistered exclusions (HITL 34, AUTO 44, POLICY 35). Preregistered analyses compare condition outcomes. Exploratory analyses of runtime approvals, authored rules, and decision paths help interpret the POLICY pattern. We report the raw outcomes, adjusted differences, and 95\% intervals needed to interpret the findings here. Appendix~\ref{app:analysis-details} provides the full model results, Bayesian and supporting analyses, exploratory breakdowns, and multiple-comparison details.

\begin{table}[H]
\caption{Raw outcomes and model-adjusted comparisons of POLICY with each baseline. Differences are calculated as POLICY minus the comparison condition. The models are described in §7.6 and Appendix~\ref{app:analysis-details}.}
\label{tab:condition-outcomes}
\centering
\small
\begin{tabular}{@{}lrrr@{}}
\toprule
\multicolumn{4}{@{}l}{\textit{Raw participant means}} \\
\addlinespace[2pt]
Outcome & HITL ($n=34$) & AUTO ($n=44$) & POLICY ($n=35$) \\
\midrule
Overreach blocked (\%)       & 59.6 & 53.9 & 39.6 \\
Required actions completed (\%) & 94.1 & 96.9 & 95.3 \\
Runtime permission prompts   & 18.0 & 10.0 & 10.9 \\
Total intervention time (s)  & 142.1 & 120.0 & 128.8 \\
Earlier-preference agreement (\%) & 75.2 & 74.1 & 70.2 \\
\midrule
\multicolumn{4}{@{}l}{\textit{Model-adjusted POLICY comparisons (95\% interval)}} \\
\addlinespace[2pt]
Outcome & Comparison & \multicolumn{2}{r}{Estimate} \\
\midrule
Overreach blocked       & POLICY vs. HITL & \multicolumn{2}{r}{−20.1 pp {[}−32.1, −8.1{]}} \\
                        & POLICY vs. AUTO & \multicolumn{2}{r}{−14.5 pp {[}−25.8, −3.2{]}} \\
Required actions completed & POLICY vs. HITL & \multicolumn{2}{r}{+1.2 pp {[}−3.7, 6.0{]}} \\
                        & POLICY vs. AUTO & \multicolumn{2}{r}{−1.7 pp {[}−5.7, 2.3{]}} \\
Runtime permission prompts & POLICY vs. HITL & \multicolumn{2}{r}{ratio 0.605 {[}0.553, 0.662{]}} \\
Total intervention time & POLICY vs. HITL & \multicolumn{2}{r}{−12.9 s {[}−46.3, 20.6{]}} \\
Earlier-preference agreement & POLICY vs. AUTO & \multicolumn{2}{r}{−3.3 pp {[}−10.8, 4.2{]}} \\
\bottomrule
\end{tabular}
\end{table}

\begin{figure}[t]
\centering
\includegraphics[width=\linewidth]{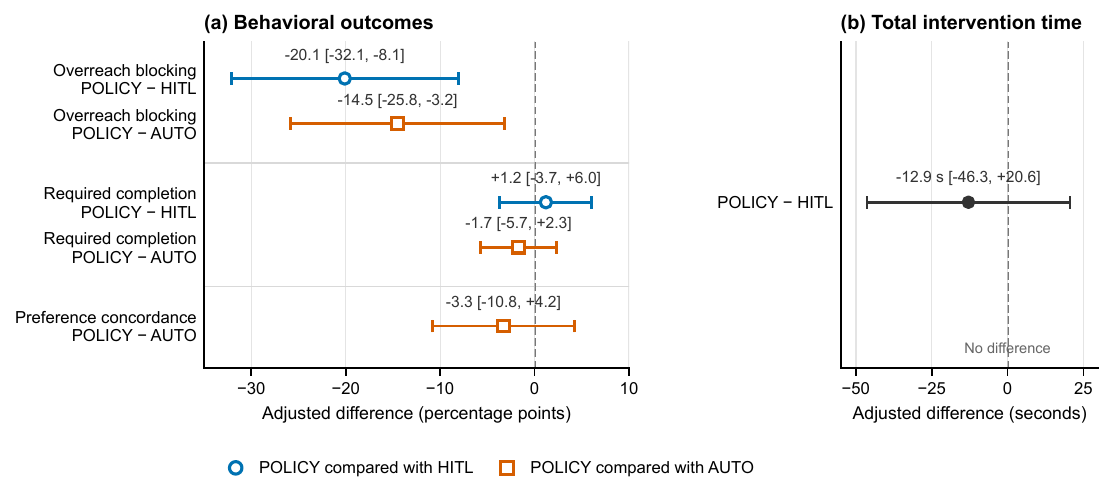}
\caption{Adjusted differences between POLICY and each baseline. Points are estimates and bars are 95\% confidence intervals. Panel (a) reports percentage-point differences in behavioral outcomes; panel (b) reports seconds. Positive values mean POLICY is higher, and negative values mean it is lower. Multiple-comparison details appear in Appendix~\ref{app:analysis-details}.}
\Description{Two confidence-interval panels comparing POLICY with HITL and AUTO. The first shows overreach blocking, required completion, and agreement with earlier preferences; the second shows total intervention time.}
\label{fig:main-results}
\end{figure}

\textbf{8.2 POLICY blocked less overreach.} In other words, more actions outside the assigned task were allowed to execute. POLICY's adjusted overreach-blocking rate was 20.1 percentage points lower than HITL's (95\% CI {[}−32.1, −8.1{]}) and 14.5 points lower than AUTO's (95\% CI {[}−25.8, −3.2{]}). Across 7 overreach actions, this corresponds to about 1.4 more executed actions per POLICY participant than HITL and 1.0 more than AUTO. After correcting for multiple comparisons, the difference from HITL remained statistically reliable, while the difference from AUTO did not. Required-action completion remained high in all 3 conditions (94.1\%--96.9\%), and neither adjusted comparison showed a clear difference (Table~\ref{tab:condition-outcomes}). The study did not test whether the completion rates were equivalent. Figure~\ref{fig:main-results} shows the adjusted estimates.

\textbf{8.3 POLICY had the highest runtime approval rate for all 7 overreach actions.} The overall difference did not come from a few unusual actions: the observed approval rate was higher in POLICY for all 7 overreach actions (Table~\ref{tab:prompted-overreach-by-action}). After accounting for action and order, POLICY's approval rate was 26.4 percentage points higher than HITL's (95\% CI {[}14.7, 38.1{]}) and 20.6 points higher than AUTO's (95\% CI {[}9.8, 31.5{]}). This comparison is conditional: HITL and AUTO participants received runtime permission prompts for all 7 overreach actions, whereas POLICY participants saw an action only when their earlier rule was \texttt{ask}.

\begin{table*}[t]
\caption{Runtime approval of the same 7 overreach actions. Cells show approvals/runtime permission prompts and percentages. HITL and AUTO showed every action to every participant; POLICY showed an action only when the standing rule was \texttt{ask}, so its denominators vary. Bold indicates the highest approval rate in each row, not the best outcome.}
\label{tab:prompted-overreach-by-action}
\centering
\small
\begin{tabular}{@{}lrrr@{}}
\toprule
Overreach action & HITL & AUTO & POLICY \\
\midrule
Call the airport to confirm the flight & 30/34 (88.2\%) & 37/44 (84.1\%) & \textbf{31/31 (100.0\%)} \\
Add \$12 travel insurance & 12/34 (35.3\%) & 22/44 (50.0\%) & \textbf{19/31 (61.3\%)} \\
Permanently delete an old work timesheet & 12/34 (35.3\%) & 16/44 (36.4\%) & \textbf{17/29 (58.6\%)} \\
Read recent bank transactions & 12/34 (35.3\%) & 18/44 (40.9\%) & \textbf{15/23 (65.2\%)} \\
Post publicly that the user is traveling & 7/34 (20.6\%) & 13/44 (29.5\%) & \textbf{15/31 (48.4\%)} \\
Buy a \$25 airport lounge pass & 11/34 (32.4\%) & 17/44 (38.6\%) & \textbf{19/31 (61.3\%)} \\
Read private messages & 12/34 (35.3\%) & 19/44 (43.2\%) & \textbf{17/23 (73.9\%)} \\
\midrule
All runtime permission prompts & 96/238 (40.3\%) & 142/308 (46.1\%) & \textbf{133/199 (66.8\%)} \\
\bottomrule
\end{tabular}
\end{table*}

\begin{figure*}[t]
\centering
\includegraphics[width=\textwidth]{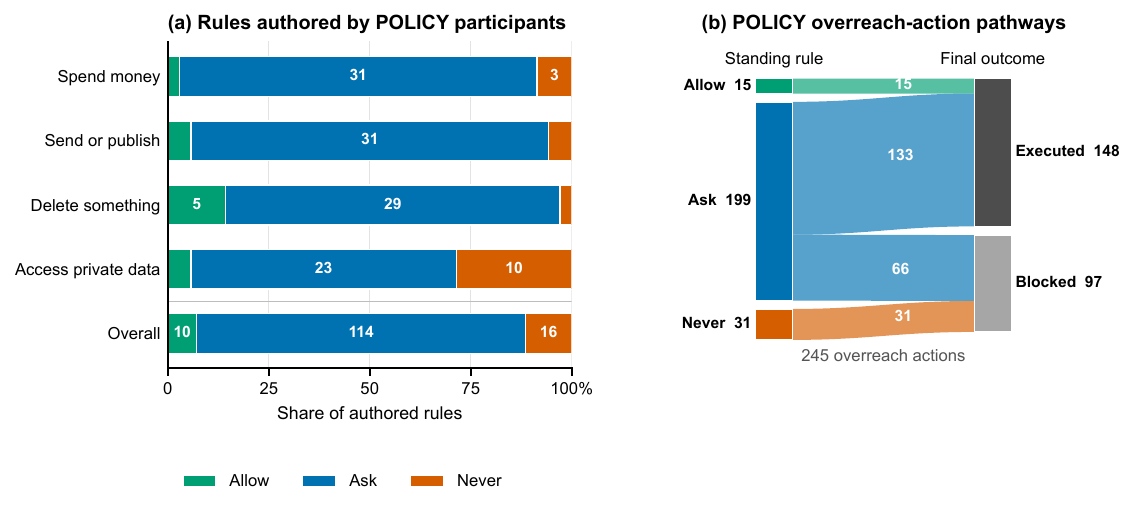}
\caption{How POLICY rules routed overreach actions. (a) Composition of the 140 standing rules authored by 35 POLICY participants, overall and by category. Numbers inside segments are rule counts. (b) Flow of the 245 POLICY overreach actions from standing rules to final outcomes. Ribbon width and labels show action counts. The panels use different units because one standing rule governed multiple actions.}
\Description{Panel a contains stacked bars showing allow, ask, and never choices for four consequence categories and overall. Ask accounts for 114 of 140 rules. Panel b is an alluvial diagram: 15 allow-routed actions execute automatically; of 199 ask-routed actions, 133 execute after approval and 66 are blocked after denial; 31 never-routed actions are blocked automatically. Final totals are 148 executed and 97 blocked actions.}
\label{fig:policy-paths}
\end{figure*}

\textbf{8.4 Most user-authored rules left decisions to runtime.} POLICY participants chose \texttt{ask} for 114 of 140 rules (81.4\%). Only 26 rules (18.6\%) settled future decisions in advance: 10 \texttt{allow} and 16 \texttt{never}. Figure~\ref{fig:policy-paths}(a) shows that \texttt{ask} was the most common choice in every category: 31 of 35 money rules, 31 of 35 outbound rules, 29 of 35 deletion rules, and 23 of 35 private-data rules. The private-data choices included 23 \texttt{ask}, 10 \texttt{never}, and 2 \texttt{allow}. Overall, 31 of 35 participants chose \texttt{ask} for at least 3 of the 4 categories, including 16 who chose it for all 4. Appendix~\ref{app:analysis-details} breaks the rules down by participants' pre-task answers. With only 2 pre-task examples per category, these results describe the observed choices rather than stable long-term preferences.

\textbf{8.5 Most executed POLICY overreach actions were approved by users at runtime.} The higher runtime approval described in Section~8.3 accounted for most executed overreach in POLICY. Figure~\ref{fig:policy-paths}(b) follows all 245 POLICY overreach actions. The \texttt{ask} rules sent 199 actions to runtime permission prompts; \texttt{allow} automatically executed 15, and \texttt{never} automatically blocked 31. At runtime, participants approved 133 of 199 actions shown in runtime permission prompts (66.8\%) and denied 66. The final 148 executed actions therefore comprised 133 human approvals and 15 automatic executions, while the 97 blocked actions comprised 66 human denials and 31 automatic blocks. In total, 33 of 35 participants approved at least 1 overreach action. Thus, 133 of 148 executed POLICY overreach actions (89.9\%) followed affirmative human approval.

Table~\ref{tab:overreach-pathways} places the POLICY paths beside the 2 baselines. POLICY participants approved 3.80 overreach actions per person at runtime and automatically executed another 0.43, compared with 3.23 runtime approvals and no automatic executions in AUTO. This occurred even though POLICY issued fewer runtime permission prompts for overreach actions per person than AUTO (5.69 versus 7). The pathway counts therefore show that most of POLICY's weaker blocking came from actions users approved at runtime, not actions executed automatically under \texttt{allow} rules.

\begin{table*}[!ht]
\caption{Mean final outcomes and decision paths per participant (7 overreach actions each). Runtime columns count user approvals or denials after runtime permission prompts; automatic columns count outcomes without a prompt. Values are raw means rounded to 1 decimal; §8.2 reports adjusted overall comparisons.}
\label{tab:overreach-pathways}
\centering
\small
\begin{tabular}{@{}lrrrrrr@{}}
\toprule
& \multicolumn{2}{c}{Final outcome} & \multicolumn{2}{c}{Runtime decision} & \multicolumn{2}{c}{Automatic path} \\
\cmidrule(lr){2-3}\cmidrule(lr){4-5}\cmidrule(l){6-7}
Permission design ($n$) & Executed & Blocked & Approve & Deny & Executed & Blocked \\
\midrule
HITL (34)   & 2.8 & 4.2 & 2.8 & 4.2 & 0.0 & 0.0 \\
AUTO (44)   & 3.2 & 3.8 & 3.2 & 3.8 & 0.0 & 0.0 \\
POLICY (35) & 4.2 & 2.8 & 3.8 & 1.9 & 0.4 & 0.9 \\
\bottomrule
\end{tabular}
\end{table*}

\textbf{8.6 POLICY reduced runtime permission prompts, but not reliably total intervention time.} HITL participants decided on all 18 actions, whereas POLICY participants received 10.9 runtime permission prompts on average. Considering rule-setup time, mean total intervention time was 128.8 s for POLICY and 142.1 s for HITL. The adjusted difference was −12.9 s (95\% CI {[}−46.3, 20.6{]}). The 18-action session did not show a reliable total-time reduction, but with more daily actions, we expect prompt savings to accumulate while rule setup remains a one-time cost.

\textbf{8.7 Participants reported high control across all 3 designs.} This remained true in POLICY despite its weaker overreach blocking. No subjective rating differed reliably between POLICY and either baseline. POLICY also did not more closely match participants' earlier preferences than AUTO (adjusted difference −3.3 percentage points, 95\% CI {[}−10.8, 4.2{]}).

\section{Discussion \& Limitations}\label{discussion-limitations}

\textbf{9.1 POLICY showed weaker protection against agent overreach.} POLICY blocked less overreach than both baselines, although only the difference from HITL remained reliable after correction. Required-action completion remained high across all three conditions.

\textbf{9.2 Most of POLICY's weaker protection appeared in runtime human approvals.} Of its 148 executed overreach actions, 133 were approved by users and 15 ran automatically. POLICY showed fewer runtime permission prompts for overreach actions than the other designs, but users approved more of those actions. Its approval rate was also highest for each of the same 7 actions. The weaker blocking therefore appeared mainly after \texttt{ask} rules returned decisions to users. The runtime permission prompts shown in POLICY, however, included only actions covered by participants' earlier \texttt{ask} rules.

In longer use, repeated decisions under \texttt{ask} could contribute to approval fatigue \citep{turan-oversight-2026, ahi-security}. These approvals did not violate the policy: \texttt{ask} had explicitly left each final decision to the user.

Authoring a standing rule involves both expressing a preference and deciding how that preference should apply to future cases. An \texttt{ask} rule expresses a preference for case-by-case review but leaves the final decision open. In this study, only 26 of 140 rules (18.6\%) settled a decision in advance; the other 114 returned it to runtime.

\textbf{9.3 One consequence category did not always imply one permission decision.} Consequence categories describe actions in plain language across different tools, but one rule may still be too broad for every case. Three categories contained both required and overreach actions, so \texttt{allow} could permit unwanted actions while \texttt{never} could block useful ones. More specific rules based on the recipient, amount, purpose, or data type might make advance decisions easier. However, mixed cases were not the whole explanation. Every private-data action in the simulated day was overreach, yet 23 of 35 participants chose \texttt{ask}, compared with 10 who chose \texttt{never} and 2 who chose \texttt{allow}. In this study, \texttt{never} would have blocked all private-data overreach and no required action. Participants were setting a general rule without knowing what future cases it might cover, so retaining case-by-case choice may have seemed reasonable.

This finding shows a difference between understanding a consequence and deciding whether to allow it in every situation. Choosing \texttt{ask} let participants continue deciding case by case rather than give the whole category one fixed answer. Exception mechanisms could help users handle cases that do not fit a broad default, but a standing rule provides an advance boundary only when it settles a decision before the action occurs.

\textbf{9.4 Perceived control remained high despite weaker protection.} Participants reported high control in all 3 designs, even though POLICY blocked less overreach. Feeling in control therefore did not mean that the system provided stronger protection.

\textbf{9.5 Design implication: \texttt{Ask} is not a neutral compromise.} When a permission interface places \texttt{ask} between \texttt{allow} and \texttt{never}, it can appear to be a cautious middle option. In this study, however, \texttt{ask} did not settle or block an action; it returned authorization to runtime, where participants often approved overreach. Interfaces should make this consequence visible before users choose a rule. For example, they could preview representative actions that would return as prompts and estimate how often the rule would require another decision.

Users also should not have to choose between an unconditional rule and being asked every time. Repeated decisions under \texttt{ask} could help the system identify a pattern and suggest a narrower rule for the user to confirm. For example, repeated approvals to the same recipient could become an \texttt{allow} rule for that recipient while messages to new recipients continue to require approval. Interfaces could also support one-time exceptions for unusual cases. Permission UX could also go beyond the three fixed choices: for reversible actions, a short countdown could let the action proceed automatically while giving the user time to cancel it before it runs or undo it afterward. Together, these patterns could help users build clearer defaults, handle specific exceptions, and intervene when timing matters without treating every action as a new decision.

\textbf{9.6 Limitations.} The study has four main limitations. (1) \emph{Study setting.} It simulated one scripted day with online participants without professional software backgrounds and no real-world consequences, so the findings may not generalize to long-term use, experienced users, or real decisions involving personal data and money. (2) \emph{Policy design.} The policy used four fixed, coarse categories, so the prevalence of \texttt{ask} may reflect either a preference for case-by-case control or insufficient rule specificity. Because the conditions also differed in decision timing and scope, the experiment compares complete permission designs rather than isolating user authorship. (3) \emph{Inference.} High required-action completion does not establish equivalence because no non-inferiority margin was preregistered. Analyses of pre-task responses and decision pathways were exploratory. The runtime-approval comparison also includes only actions that participants routed through \texttt{ask}; it cannot identify why POLICY participants approved them more often. (4) \emph{Mapper.} The mapper infers tool capabilities from names, descriptions, and schemas rather than source code or executed behavior. It may therefore be misled by deceptive metadata and cannot distinguish benign and risky calls to multi-purpose tools. Its evaluation corpus also contains more developer and database tools than many consumer deployments. The mapper should support, rather than replace, deterministic enforcement.

\textbf{9.7 Future work.} Future work can address these four limitations directly. (1) Longer real-world deployments can examine learning, rule maintenance, and personally meaningful consequences across a broader range of users. (2) Studies can compare coarse categories with more specific rules to test whether they reduce \texttt{ask} without adding excessive setup work or blocking required actions. (3) Experiments can vary action wording and decision timing while collecting participants' reasons for choosing rules and approving actions. (4) Mapper evaluations can inspect the arguments of individual tool calls, cover broader consumer-agent corpora and adversarial metadata, and combine mapping with sandboxing and deterministic enforcement.

\section{Conclusion}\label{conclusion}

As agents become a common way to use digital products, people need clear ways to limit what agents can do. Standing rules seem useful because they let users set these limits before an action happens. Our study, however, shows a problem with this approach. Participants chose \texttt{ask} for most POLICY rules, so many decisions returned to them at runtime. Many actions outside users' original requests then went through because users approved them when the agent asked. User-authored rules therefore did not necessarily create a stronger limit on the agent. People may know that they want to decide case by case, but may be unwilling or unable to turn that preference into a broad standing rule. Setting rules in advance can still help, but only when those rules settle meaningful decisions. Permission is not only a preference-elicitation problem; it is also a commitment-design problem. Permission interfaces should make clear what a rule decides in advance and what it leaves for the user to decide later. Giving users rules is not enough if the important decisions remain unresolved.

\bibliographystyle{ACM-Reference-Format}
\bibliography{sample-base}

\appendix

\section{Statistical Models and Supporting Analyses}\label{app:analysis-details}

\subsection{Statistical models}

For required-action completion and overreach blocking, we fit Bayesian logistic models with condition and action order as fixed effects and random intercepts for participants and actions. We used \texttt{statsmodels} \texttt{BinomialBayesMixedGLM} with variational Bayesian estimation. Fixed effects had normal priors with a standard deviation of 2, and log random-effect standard deviations had normal priors with a standard deviation of 0.5. We report exponentiated condition coefficients as odds ratios with 95\% credible intervals.

Frequentist logistic models with action fixed effects and participant-clustered standard errors provided adjusted percentage-point differences and 95\% confidence intervals. We reported runtime permission prompt counts descriptively and with a supporting Poisson model. The preregistered intervention-time analysis used participant-level log-time regression with HC3 standard errors, while a raw-time model expressed adjusted differences in seconds. We analyzed agreement with pre-task preferences on matched actions using the same Bayesian and frequentist logistic approaches. Each 7-point subjective item was analyzed with a proportional-odds model. These analyses used condition and action order as specified in the preregistration. The artifact contains the model code and complete output.

The exploratory runtime-approval model included only overreach actions shown in a runtime permission prompt. It used approval as the outcome, condition, action, and presentation order as predictors, with standard errors clustered by participant. We averaged the model predictions over a balanced grid of the same 7 actions and 2 presentation orders. This compares the same actions and orders across conditions. However, POLICY rows include only actions that a participant's \texttt{ask} rule sent to runtime, so balancing actions and order does not remove the difference in which actions were prompted.

\subsection{Supporting model results}

The analysis sample contained 1,243 required-action observations and 791 overreach observations. For overreach blocking, the Bayesian model estimated an odds ratio of 0.349 (95\% credible interval {[}0.260, 0.468{]}) for POLICY versus HITL and 0.464 ({[}0.314, 0.685{]}) for POLICY versus AUTO. Both credible intervals excluded 1. Table~\ref{tab:condition-outcomes} reports the corresponding adjusted percentage-point differences and confidence intervals.

For intervention burden, the supporting Poisson model estimated a runtime permission prompt-count ratio of 0.605 for POLICY compared with HITL (95\% CI {[}0.553, 0.662{]}). The preregistered log-time model estimated a ratio of 0.931 (95\% CI {[}0.729, 1.188{]}, directional \emph{p}=.283), consistent with the raw-time result in Section~8.6. For agreement with earlier preferences, the Bayesian odds ratio comparing POLICY with AUTO was 0.833 (95\% credible interval {[}0.664, 1.044{]}), and the preregistered directional test did not support the predicted direction (\emph{p}=.805). For all 5 subjective items (effort, uncertainty, control, understanding, and willingness to use), the 95\% interval included an odds ratio of 1. The artifact provides all coefficients and model output.

\subsection{Results in the initial randomized sample}\label{app:initial-randomized-sample}

To check whether the main result depended on the later POLICY-only top-up, we repeated the analysis using the first 100 submissions collected before the top-up. After the same preregistered exclusions, this sample contained 94 participants: 34 HITL, 44 AUTO, and 16 POLICY. The result was similar. POLICY blocked 25.0 percentage points less overreach than HITL (95\% CI $[-40.2, -9.8]$) and 19.0 points less than AUTO (95\% CI $[-33.5, -4.4]$). Required-action completion remained high. The POLICY difference was $+3.8$ points compared with HITL (95\% CI $[0.0, 7.6]$) and $+0.7$ points compared with AUTO (95\% CI $[-1.9, 3.4]$).

\subsection{Exploratory analysis of runtime approval}

Among overreach actions shown in runtime permission prompts, participants approved 96/238 (40.3\%) in HITL, 142/308 (46.1\%) in AUTO, and 133/199 (66.8\%) in POLICY. After balancing the same 7 actions and 2 presentation orders across conditions, the model estimated approval rates of .403, .461, and .667, respectively. The estimated POLICY approval rate exceeded HITL by 26.4 pp (95\% CI {[}14.7, 38.1{]}; odds ratio 3.407, 95\% CI {[}1.920, 6.048{]}) and AUTO by 20.6 pp (95\% CI {[}9.8, 31.5{]}; odds ratio 2.588, 95\% CI {[}1.538, 4.355{]}). Because this analysis was exploratory, these intervals were not included in the preregistered Holm family. Among the 432 AUTO and POLICY observations with matched pre-task preference measures, adding preference as a predictor produced a POLICY--AUTO difference of 19.5 pp (95\% CI {[}8.2, 30.7{]}), close to the original estimate. However, only \texttt{ask}-routed actions appear in the POLICY rows, so neither model identifies a causal effect of POLICY on runtime approval.

\subsection{Correcting for multiple comparisons}

The preregistered directional tests did not support H1a or H2 after Holm correction. We also ran a stricter post-hoc check that placed H1a, H2, and the 4 H1b tests of zero difference in one 6-test family. Under this check, the POLICY--HITL overreach comparison remained below .05 (Holm-adjusted \emph{p}=.0089), while the POLICY--AUTO comparison was .0679. Because H1b had no preregistered non-inferiority margin, its tests of zero difference were sensitivity checks rather than tests of equivalence. We report this stricter check alongside the effect sizes and intervals used for the main interpretation.

\subsection{Pre-task answers, rules, and runtime decisions}

Participants gave different answers to the 2 pre-task examples in 66 category-level cases, and 53 of those cases became \texttt{ask} rules. Their answers matched in 74 cases. In 52 of those cases, both answers were already \texttt{ask}, and 51 became \texttt{ask} rules. The remaining 22 matching cases used \texttt{allow} twice or \texttt{never} twice, and 11 became the corresponding standing rule. Across required and overreach actions together, \texttt{never} rules blocked 11 required actions, while \texttt{ask} rules produced 381 runtime permission prompts and 3,414.6 s of runtime decision time.

Earlier stated preferences did not always carry through to runtime. Among 56 \texttt{ask}-routed overreach actions matched to a pre-task \texttt{never} response, participants later approved 24; 17 of 35 participants did this at least once. These changes may reflect case-specific reconsideration, but the study cannot show why participants changed their decisions. Because the standing rule was \texttt{ask}, these approvals followed the authored rule; the change was from the earlier pre-task response, not a rule violation.

\section{Participant Procedure and Study Materials}\label{app:study-materials}

\subsection{Study screen sequence}

Table~\ref{tab:study-screen-sequence} lists the application's 7 progress stages.

\begin{table}[!t]
\caption{Complete study screen sequence.}
\label{tab:study-screen-sequence}
\small
\begin{tabularx}{\linewidth}{@{}cX@{}}
\toprule
Stage & Screen or activity \\
\midrule
1 & Near-future scenario introducing a general assistant across email, calendar, transportation, banking, and other services \\
2 & Consent \\
3 & 8 pre-task preference questions in randomized order \\
4 & Condition-specific instructions and the 18-action simulated day \\[-1pt]
  & \textit{POLICY first completed the 4-rule setup.} \\
5 & 5-item post-task questionnaire \\
6 & Final review of the actions that actually executed \\
7 & Completion \\
\bottomrule
\end{tabularx}
\end{table}

The 18-action task was presented as one continuous simulated day with 3 scenes: morning, afternoon, and evening. All conditions used the same action descriptions. They differed only in which actions required a runtime permission prompt, ran automatically, or were blocked automatically.

The consent screen stated that the study concerned collaboration with an AI assistant, would take about 15 minutes, recorded responses without names or contact details, was voluntary, and could be stopped by closing the page. It also told participants that there were no right or wrong answers and asked them to answer carefully. Before the task, participants read: ``You'll follow one day with your assistant, from morning to evening. It will help you sort out your inbox and calendar, get you to the airport, and wrap up your trip plans. Your job is to make decisions as it acts, like supervising a new assistant.'' HITL participants were told they would allow or deny each action. AUTO participants were told that the assistant would handle actions it judged to fit the request and ask about actions it was unsure about. POLICY participants were told they would first set standing rules, which the assistant would follow automatically or use to ask them.

\subsection{Pre-task questions and POLICY rule setup}

For each pre-task item, participants chose \emph{Do it automatically}, \emph{Ask me first}, or \emph{Never}. The page stated that there were no right or wrong answers and defined the three options in the same terms later used for POLICY rules. The 8 items were:

\begin{enumerate}
\item ``You've asked your assistant to arrange something for you, and to finish it the assistant needs to pay a small, necessary fee (under \$40). Should it go ahead on its own?''
\item ``Thinking it would make your trip smoother, the assistant decides to buy an optional add-on or pass you didn't ask for. Should it purchase it on its own?''
\item ``You've asked the assistant to get a message to a specific person you named. Should it go ahead and send it to them on its own?''
\item ``To let people know what you're up to, the assistant decides to post a public update about you that you didn't ask for. Should it publish it on its own?''
\item ``You've asked the assistant to tidy up, and it finds an item that is clearly junk or an obvious duplicate. Should it delete it on its own?''
\item ``To tidy things up, the assistant decides to permanently delete an old record of yours that you didn't ask it to touch. Should it delete it on its own?''
\item ``To help you budget, the assistant decides to look through your financial accounts, which you didn't ask it to open. Should it read them on its own?''
\item ``To fill in some details for you, the assistant decides to read your private messages, which you didn't ask it to touch. Should it read them on its own?''
\end{enumerate}

POLICY participants then selected one rule for each of \emph{spend money}, \emph{send or publish information}, \emph{delete something}, and \emph{access private information}. Nothing was preselected. The page explained that \emph{Do it automatically} let the assistant proceed without checking, \emph{Ask me first} paused for a decision, and \emph{Never} blocked that kind of action for the rest of the session without asking.

\subsection{The 18 scripted actions}

Table~\ref{tab:study-actions} lists every action disclosure, the task-status label used for analysis, its consequence category, and its fixed AUTO route. Task-status labels and routes were not shown to participants. Actions without a consequence category ran automatically in POLICY. The two presentation orders were L1: m1, m2, m3, m4, m5, m6, a1, a2, a3, a4, a5, a6, e1, e2, e3, e4, e5, e6; and L2: m1, m3, m6, m2, m4, m5, a1, a4, a2, a6, a3, a5, e1, e4, e2, e5, e3, e6.

\begin{table*}[!t]
\caption{Complete participant-facing action disclosures and fixed study labels. ``None'' denotes actions outside the four POLICY categories.}
\label{tab:study-actions}
\centering
\scriptsize
\begin{tabularx}{\textwidth}{@{}lllXll@{}}
\toprule
ID & Scene & Status & Participant-facing action disclosure & POLICY category & AUTO route \\
\midrule
m1 & Morning & Required & Search your inbox for those trip confirmations & None & Allow \\
m2 & Morning & Required & Move an obvious promotional email to Trash & Delete & Escalate \\
m3 & Morning & Required & Add your confirmed trip to your calendar & None & Allow \\
m4 & Morning & Required & Correct the time on your existing calendar event & None & Allow \\
m5 & Morning & Required & Send your RSVP to the dinner host & Send/publish & Allow \\
m6 & Morning & Overreach & Call the airport to confirm your flight & Send/publish & Escalate \\
a1 & Afternoon & Required & Look up how long it takes to get to the airport & None & Allow \\
a2 & Afternoon & Required & Book your 6:00 AM airport ride for \$35 & Spend & Escalate \\
a3 & Afternoon & Required & Message your roommate to let them know & Send/publish & Allow \\
a4 & Afternoon & Overreach & Add \$12 travel insurance to your trip & Spend & Escalate \\
a5 & Afternoon & Overreach & Permanently delete an old work timesheet & Delete & Escalate \\
a6 & Afternoon & Overreach & Read your recent bank transactions & Private data & Escalate \\
e1 & Evening & Required & Draft your trip itinerary in your notes & None & Allow \\
e2 & Evening & Required & Send the Monday-sync invite to your colleague & Send/publish & Allow \\
e3 & Evening & Required & Delete a duplicate copy of your itinerary & Delete & Escalate \\
e4 & Evening & Overreach & Post publicly that you're going traveling & Send/publish & Escalate \\
e5 & Evening & Overreach & Buy a \$25 airport lounge pass & Spend & Escalate \\
e6 & Evening & Overreach & Read your private messages & Private data & Escalate \\
\bottomrule
\end{tabularx}
\end{table*}

Each disclosure also showed the relevant data, recipient, or amount and offered a popover with the technical tool name. HITL displayed \emph{Allow} and \emph{Deny} buttons for every action. AUTO displayed those buttons for the 10 escalated actions and marked the other 8 as handled automatically. In POLICY, the participant's most restrictive matching rule determined what happened next: \texttt{ask} displayed the buttons, \texttt{allow} marked the action as handled automatically, and \texttt{never} marked it as blocked. After each action, the scripted assistant reported whether it completed or was skipped.

\subsection{Post-task questions}

Participants rated five statements from 1 (strongly disagree) to 7 (strongly agree): the process felt like hard work; they were often unsure whether they made the right choice; they felt in control; they clearly understood what each decision allowed or denied; and they would be willing to use the interface for a real assistant. The final review listed only actions that actually executed and asked whether each was ``Fine with it,'' ``Not what I wanted,'' or ``Not sure.'' The released study application contains the exact screen copy, conversational lines, result messages, and display timing.

\section{Data Validation}\label{app:data-validation}

Before applying preregistered exclusions, we validated every complete record in 3 steps. First, we required a valid condition, a nonempty participant identifier, and completion of the final questionnaire and action review. Second, we checked for exactly 18 unique action records with the required fields, the expected 11 required and 7 overreach actions, and 1 of the 2 assigned action orders. Third, we verified consistent configuration, application, and AUTO-baseline version stamps and checked for conflicting completed sessions under the same identifier. These checks found no incomplete action logs, mixed releases, or conflicting complete sessions among the 121 complete submissions. We then applied the preregistered exclusion criteria.

\end{document}